\documentclass[graybox]{svmult}
\usepackage{type1cm}        
\usepackage{makeidx}         
\usepackage{graphicx}        
\usepackage{multicol}        
\usepackage[bottom]{footmisc}
\usepackage{newtxtext}       %
\usepackage{epstopdf}       %
\usepackage{newtxmath}
\usepackage{geometry}
\makeindex             

\begin{document}

\title*{Production of electrons from heavy-flavour hadron decays in different collision systems with ALICE at LHC}

\author{Sudhir Pandurang Rode (for ALICE collaboration)}

\institute{Indian Institute of Technology  Indore, Indore - 453552, Madhya Pradesh, India,\\ \email{sudhir.pandurang.rode@cern.ch}}

\maketitle

\abstract{Heavy-flavour quarks, due to their large masses, are produced in the early stages of the relativistic heavy-ion collisions via initial hard scatterings. Therefore, as they experience the full system evolution, heavy quarks are effective probes of the hot and dense medium created in such collisions. In pp collisions, the measurement of heavy-flavour hadron production cross sections can be used to test our understanding of the Quantum ChromoDynamics (QCD) in the perturbative regime. Also, pp collisions provide a crucial reference for the corresponding measurements in larger systems. In Pb--Pb (Xe--Xe) collisions, the measurement of the nuclear modification factor of heavy-flavour hadrons provides information on the modification of the invariant yield with respect to pp collisions due to the produced cold and hot QCD matter. The possible mass dependence of the parton energy loss can be studied by comparing the $R_{\rm AA}$ of pions, charm and beauty hadrons. In this contribution, recent results from ALICE at the LHC are reported with focus on the different measurements of the heavy-flavour electrons in pp collisions at 2.76, 5.02, 7 and 13 TeV and in Pb--Pb (Xe--Xe) collisions at 5.02 (5.44) TeV. The results include the differential production cross sections and nuclear modification factors of heavy-flavour electrons at mid-rapidity. The comparison of experimental data with model predictions is discussed.}

\section{Introduction}

A Large Ion Collider Experiment (ALICE) is one of the four major experiments carried out at the LHC at CERN. It is a dedicated heavy-ion experiment aiming at the study of the strongly interacting matter consisting of thermally equilibrated deconfined quark and gluons, also called as Quark--Gluon Plasma. Heavy-flavour quarks are among the most important probes to understand the nature of the QGP. Due to their large masses ($m_{\rm Q}$ $\gg$ $\Lambda_{\rm QCD}$), they are produced at the early stage of heavy-ion collisions via hard scattering making them crucial probes since they witness the full evolution of the Quantum ChromoDynamics (QCD) medium. Their production mechanism can be described theoretically by perturbative QCD in the full transverse momentum range \cite{Averbeck:2013oga}. The heavy-flavour hadrons can be measured via electrons originating from their semi-leptonic decay channel.

\section{Heavy-Flavour electron measurement with ALICE}

Measurements of electrons from heavy-flavours hadron decays require excellent particle identification capabilities and reconstruction efficiencies, which are provided by the ALICE detector. Charged particles are identified using the Inner Tracking System (ITS), Time Projection Chamber (TPC) and Time of Flight (TOF) detectors using specific energy loss and time of flight information of the particles. A more detailed description of the ALICE detector can be found here \cite{Aamodt:2008zz}. The contribution of heavy-flavour electrons to the inclusive electron distribution is estimated after eliminating the primary source of background, i.e. electrons from Dalitz decays and photon conversions using electron-positron pair selections \cite{Acharya:2018upq}.

In the case of electrons from beauty hadrons decays, the distribution of the distance of closest approach (DCA) of the electrons to the primary vertex are used. A larger DCA distribution of the signal compared to the background allows their separation. DCA templates of electrons from different sources are obtained from Monte Carlo (MC) simulations. The DCA distribution of the electrons from the various sources are fitted to the DCA distribution of inclusive electrons in the data using a maximum likelihood fit method.

\section{Results and discussions}
In this section, recent results on the production cross-section of electrons coming from heavy-flavour hadron decays and their nuclear modification factors ($R_{\rm AA}$) in Pb--Pb and Xe--Xe collisions at various center-of-mass energies are shown.
\subsection{Invariant cross-sections in proton-proton collisions}
\begin{figure}[h!]
\begin{center}
\includegraphics[scale=0.45]{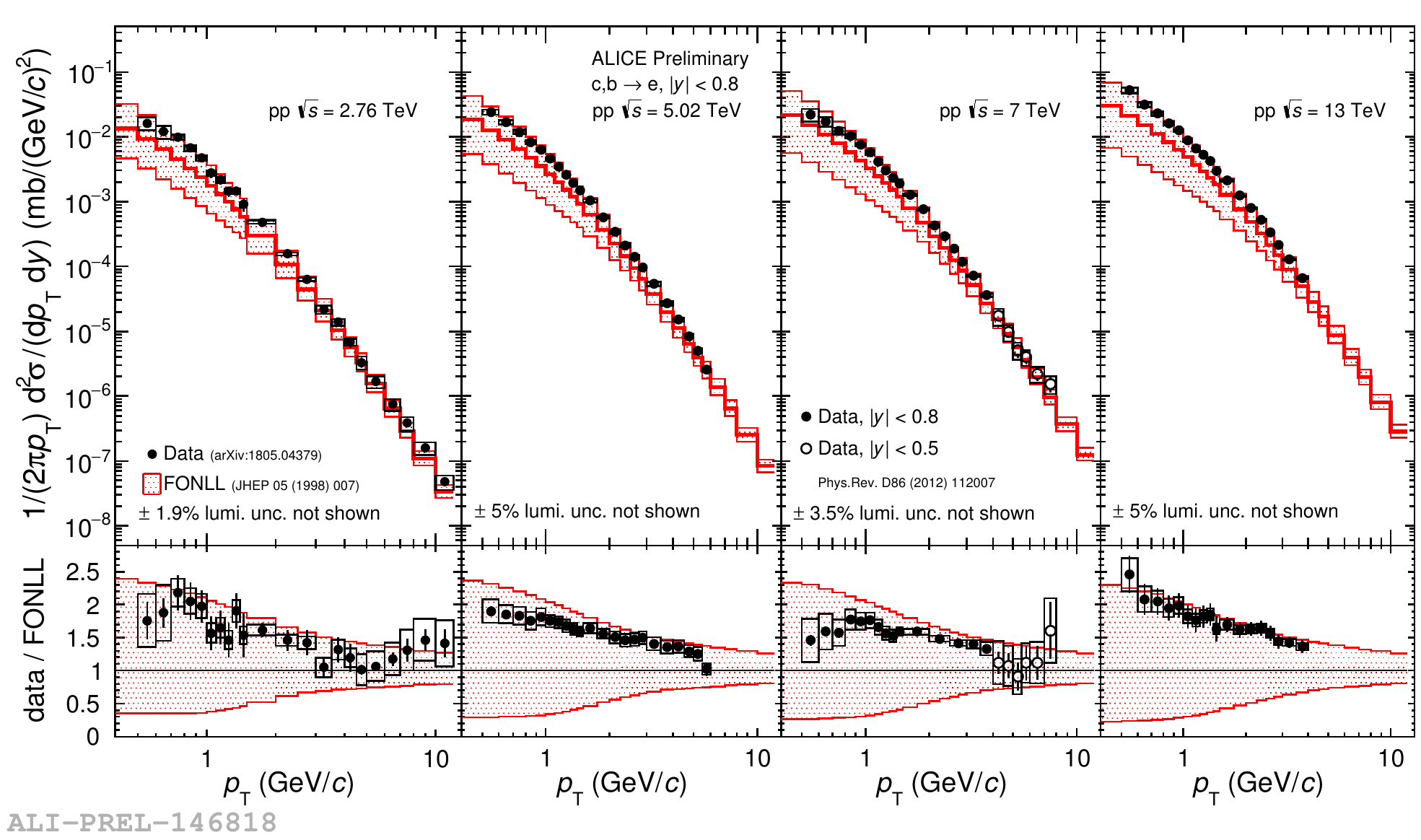}\\
\includegraphics[scale=0.45]{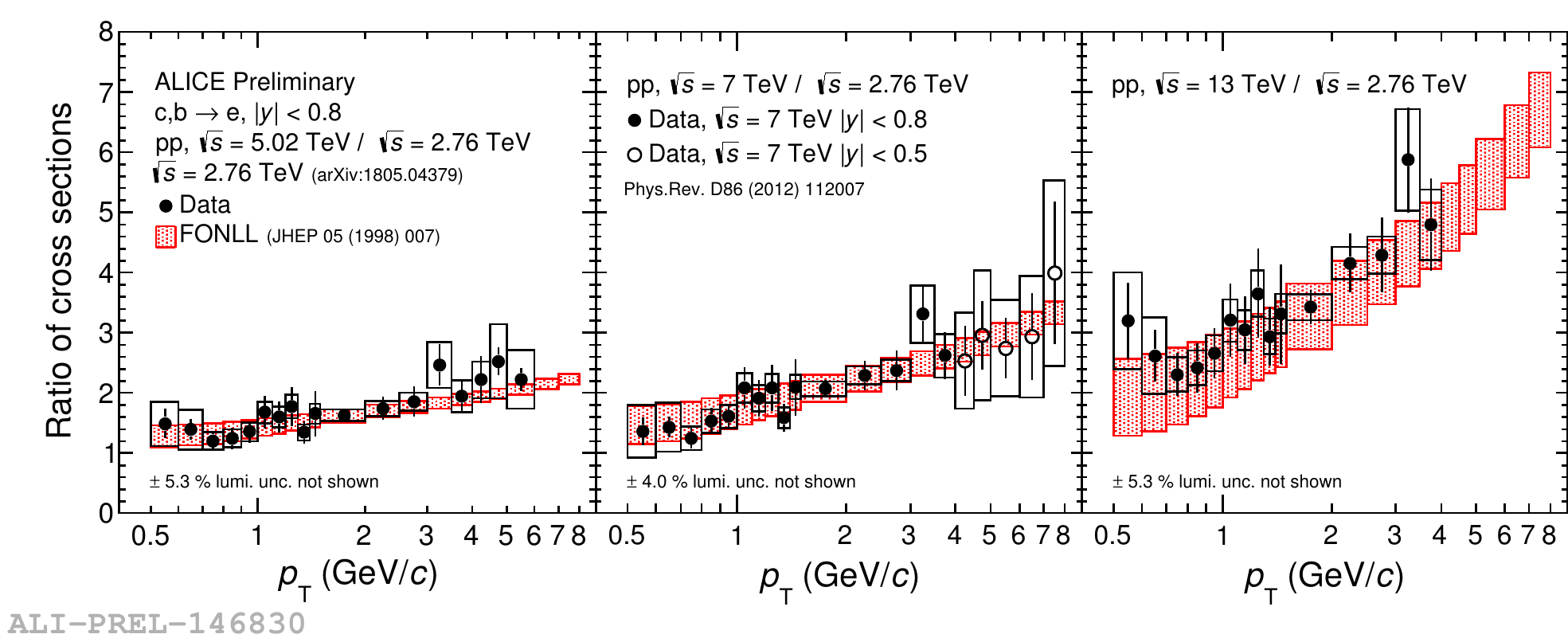}
\caption{$p_{\rm T}$-differential cross-sections of heavy-flavour electrons in pp collisions at different colliding energies.}
\label{pp}
\end{center}

\end{figure}
The $p_{\rm T}$-differential production cross-sections of heavy-flavour electrons are measured using the so-called data-driven method, i.e. photonic-electron tagging method at $\sqrt{s}$ $=$ 2.76, 5.02, 7 and 13 TeV. At 7 TeV, the data-driven method allows a reduction by a factor 3 of the systematic uncertainties in comparison to previous publications \cite{Abelev:2012xe}. As shown in Fig. \ref{pp} (upper), all the measured cross-sections are in agreement with FONLL prediction \cite{Cacciari:1998it}.  The theoretical uncertainties are reduced when forming the ratio between the cross sections at different energies, hence the ratios of the measured cross sections are shown in the lower panel of Fig. \ref{pp} and are found to be in agreement with FONLL predictions.

\subsection{$R_{\rm AA}$ of electrons from heavy-flavour decays in Pb--Pb and Xe--Xe collisions}

\begin{figure}[h!]
\begin{center}
\includegraphics[scale=0.26]{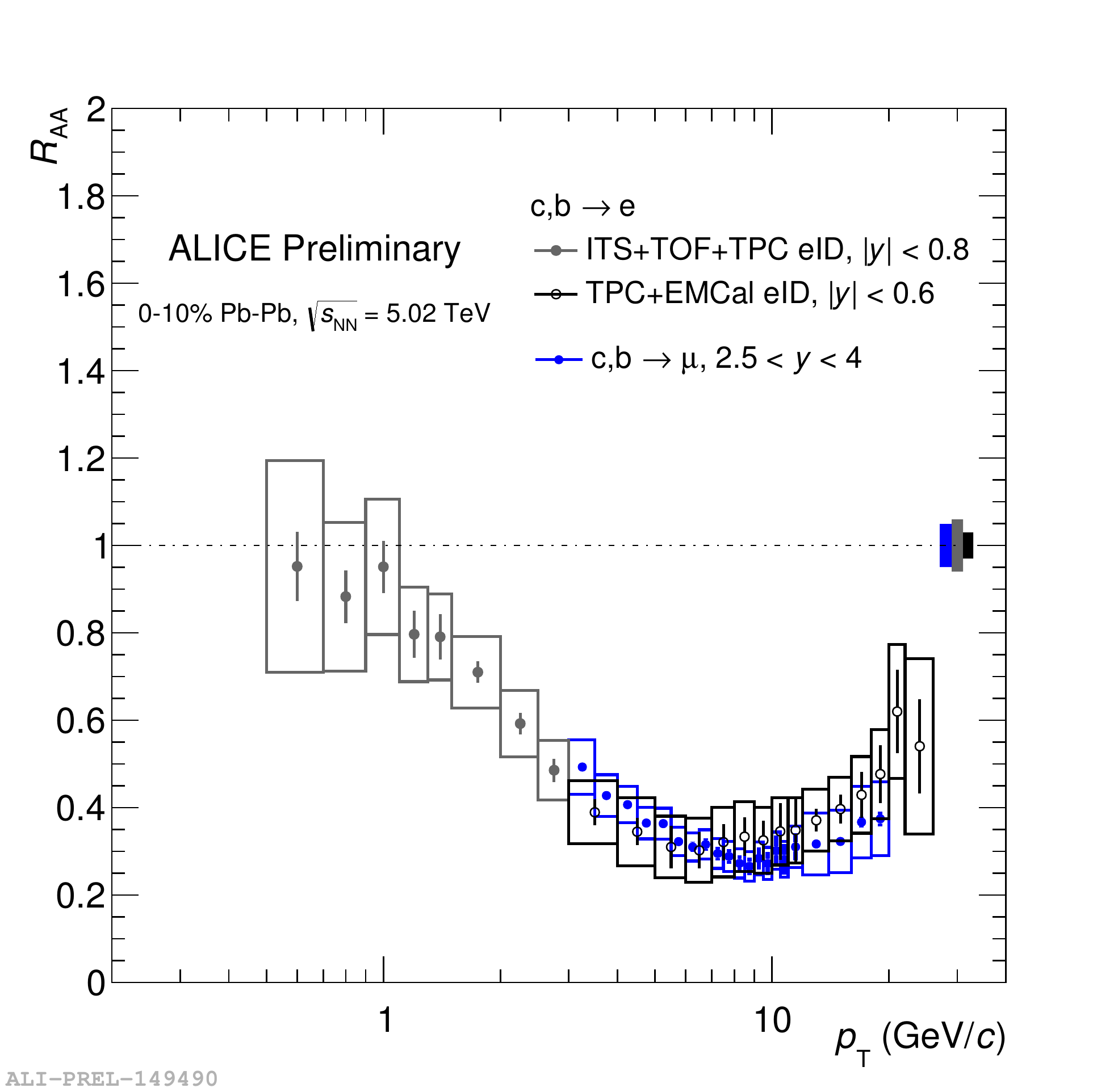}
\includegraphics[scale=0.26]{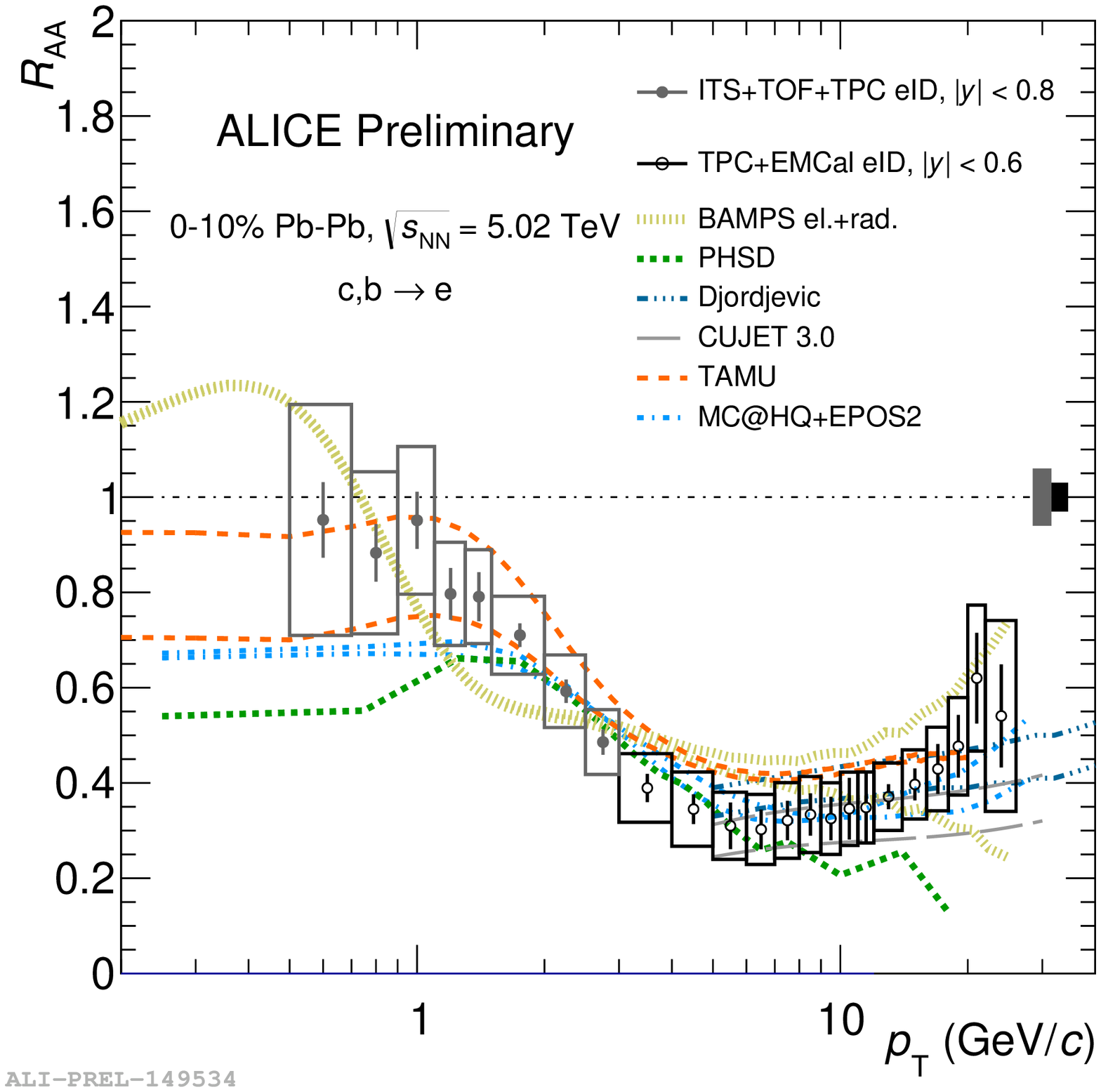}
\caption{Comparison of $R_{\rm AA}$ of heavy-flavour electrons in Pb--Pb collisions at $\sqrt{s}$ $=$ 5.02 TeV with $R_{\rm AA}$ of heavy-flavour muons (left) and with theoretical models  (right).}
\label{pbpb}
\end{center}
\end{figure}
\begin{figure}[h!]
\begin{center}
\includegraphics[scale=0.5]{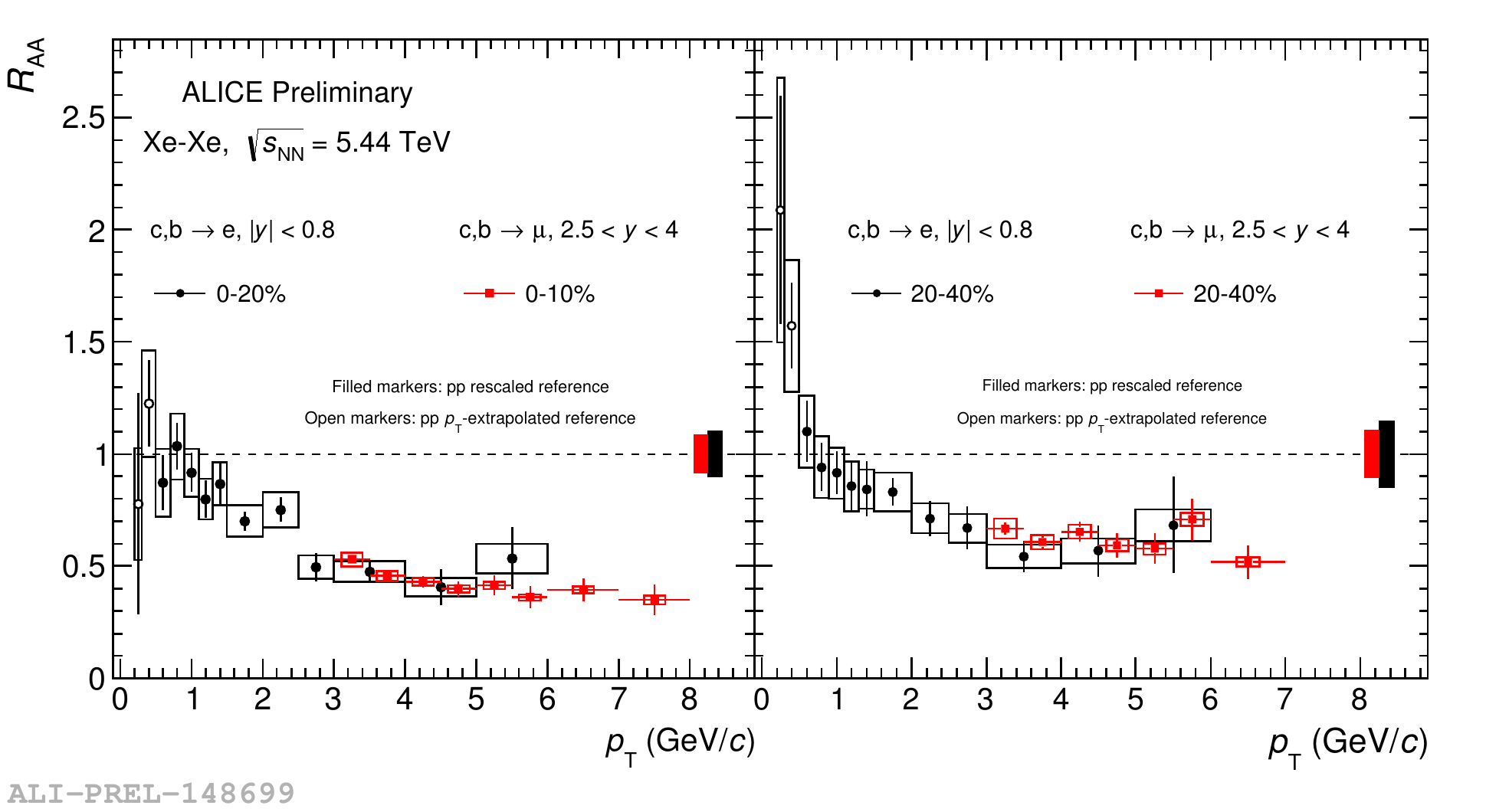}
\caption{Comparison of $R_{\rm AA}$ of electrons from heavy-flavour hadron decays in Xe--Xe collisions with $R_{\rm AA}$ of muons from heavy-flavour hadron decays.}
\label{xexe}
\end{center}
\end{figure}
The $R_{\rm AA}$ of heavy-flavour electrons in Pb--Pb and Xe--Xe collisions at $\sqrt{s_{NN}}$ $=$ 5.02 and 5.44 TeV, respectively, are measured to study the energy loss of heavy quarks inside the QGP medium. Similarly to the analysis in pp collisions, the contribution from the heavy-flavour electrons is estimated using the photonic-electron tagging method.

The $R_{\rm AA}$ of heavy-flavour electrons is shown in Fig. \ref{pbpb} and is found to be compatible with the $R_{\rm AA}$ of heavy-flavour decay muons measured in central (0--10$\%$) Pb--Pb collisions at $\sqrt{s_{NN}}$ $=$ 5.02 TeV at forward rapidity (2.5 $<$ y $<$ 4). In the very low $p_{\rm T}$ region, the measured $R_{\rm AA}$ is compatible with the TAMU \cite{He:2014cla} prediction. At higher $p_{\rm T}$, the measurement agrees with others models (Djordjevic \cite{Djordjevic:2015hra}, CUJET 3.0 \cite{Xu:2015bbz})

In Xe--Xe, the pp reference for $R_{\rm AA}$ estimation is obtained by interpolating the pp references measured at $\sqrt{s}$ $=$ 5.02 and 7 TeV  \cite{Acharya:2018eaq}. The $R_{\rm AA}$ of heavy-flavour electrons is compared with the $R_{\rm AA}$ of heavy-flavour muons. Similar suppression is observed for the muons in both centrality classes as shown in Fig. \ref{xexe}.

\subsection{$R_{\rm AA}$ of electrons from beauty quark decays in Pb--Pb collisions}
\begin{figure}[h!]
\begin{center}

\includegraphics[scale=0.3]{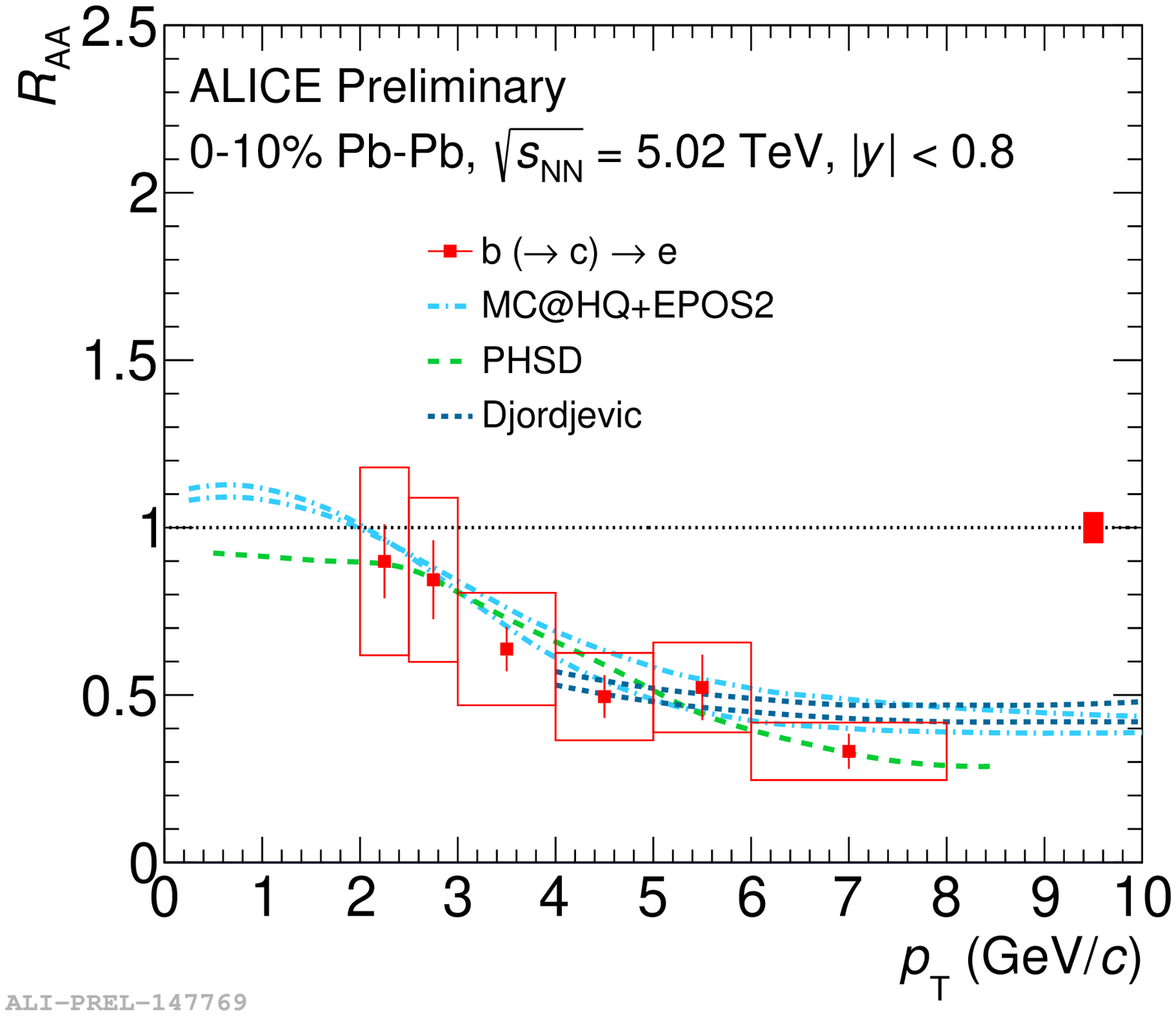}
\includegraphics[scale=0.3]{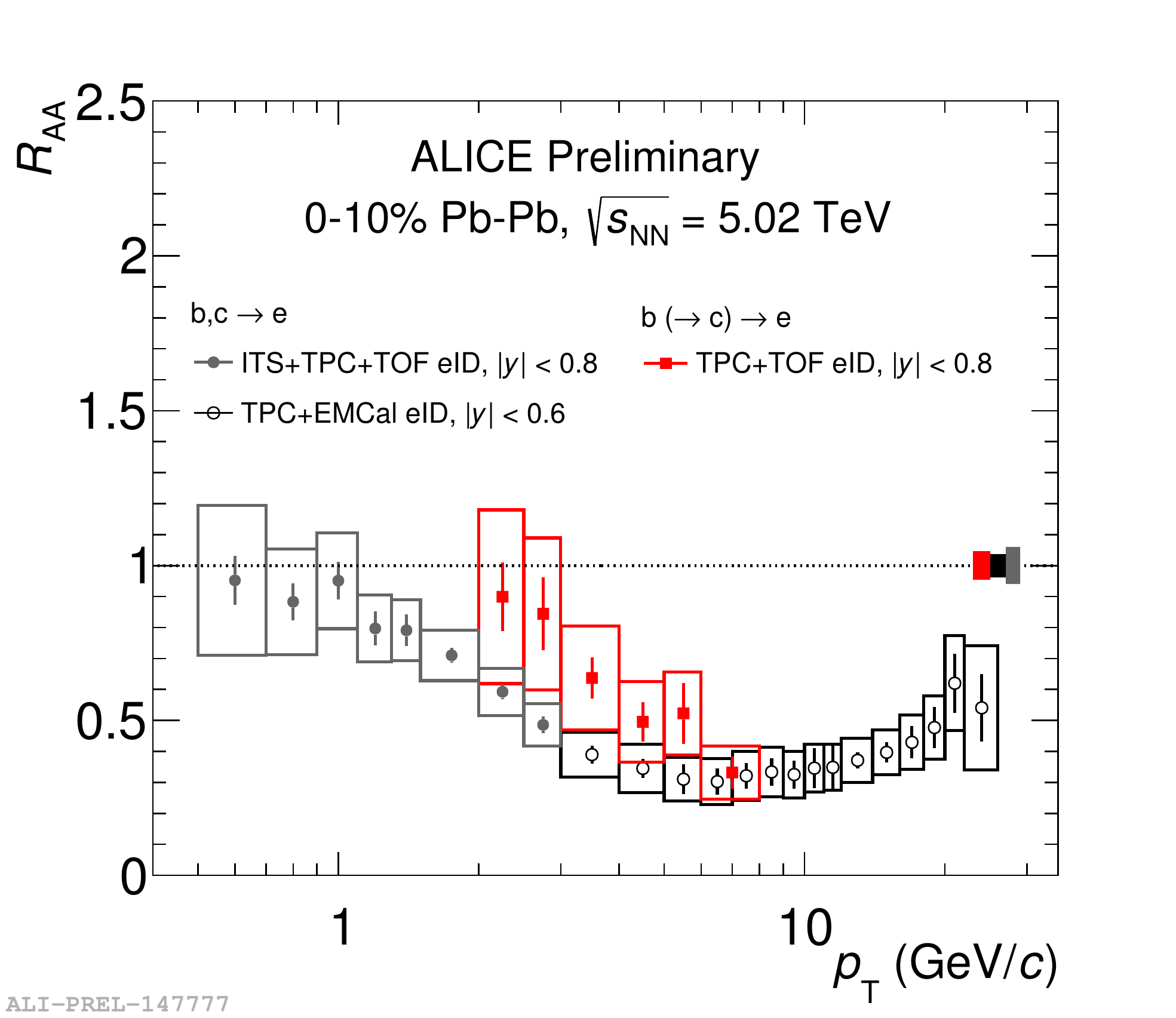}
\caption{Comparison of $R_{\rm AA}$ of electrons coming from beauty hadron decays in Pb--Pb collisions at $\sqrt{s}$ $=$ 5.02 TeV}

\label{beautypbpb}
\end{center}
\end{figure}

The $R_{\rm AA}$ of electrons from beauty hadron decays in 0--10$\%$ central Pb--Pb collisions at $\sqrt{s_{NN}}$ $=$ 5.02 TeV is shown in Fig. \ref{beautypbpb}. The yield of electrons from beauty hadron decays is estimated using the DCA template fit method. 

As shown in Fig. \ref{beautypbpb} (right), a smaller suppression of electrons from beauty quark decays with respect to electrons from heavy-flavour decays hints to a mass dependence of the quark energy loss in the QGP medium. This measurement is consistent with models that consider mass-dependent radiative and collisional energy losses \cite{Nahrgang:2013xaa, Bratkovskaya:2011wp}. For $R_{\rm AA}$ calculation,  the pp reference is obtained by an energy interpolation from 7 to 5 TeV using FONLL. Data analysis of new pp reference at $\sqrt{s}$ $=$ 5 TeV is ongoing, which would reduce the systematic and statistical uncertainties in the $R_{\rm AA}$ measurement and will give more precise results.

\section{Summary}

In summary, we have presented an overview of the measurements of electrons from the decay of heavy-flavour hadrons in pp, Pb--Pb (Xe--Xe) collisions at different energies. The $p_{\rm T}$-differential production cross-sections of heavy-flavour electrons in pp collisions at 2.76, 5.02, 7 and 13 TeV are measured and are in agreement with FONLL predictions. The nuclear modification factor of the heavy-flavour electrons is measured in Pb--Pb and Xe--Xe collisions at 5.02 and 5.44 TeV, respectively, and is found to be consistent with the $R_{\rm AA}$ of heavy-flavour muons at forward rapidity (2.5 $<$ y $<$ 4). The $R_{\rm AA}$ of beauty decay electrons, while compatible with the one of heavy-flavour electrons, hints at a separation at low $p_{\rm T}$ that would points toward the mass-dependent energy loss of the quarks inside the medium. It is also consistent with theoretical predictions, which include collisional and radiative energy losses.

With the ongoing detector upgrades \cite{Abelevetal:2014cna}, the precision on the measurement will considerably increase. The improved impact-parameter resolution, together with the improved luminosity of the LHC accelerator complex, will improve the significance of the upcoming measurements.

\end{document}